\documentstyle[psfig]{mn}

\title[A Parkes half-Jansky sample of GPS galaxies]{A Parkes half-Jansky sample of GPS galaxies}

\author[I.A.G. Snellen et al.]{I.A.G. Snellen$^{1}$, M.D. Lehnert$^2$, M.N. Bremer$^3$,
R.T. Schilizzi$^{4,5}$\\ 
$^{1}$ Institute for Astronomy, Blackford Hill, Edinburgh EH9 3HJ, United Kingdom\\
$^{2}$ Max-Planck-Institut f\"ur extraterrestrische Physik (MPE), Postfach 1312, 85741 Garching, Germany\\
$^{3}$ Department of Physics, Bristol University, H H Wills Physics Laboratory, Tyndall Avenue, Bristol, BS8 1TL, United Kingdom\\
$^{4}$Joint Institute for VLBI in Europe, Postbus 2, 7990 AA, Dwingeloo, The Netherlands\\
$^{5}$Leiden Observatory, P.O. Box 9513, 2300 RA, Leiden, The Netherlands \\
}

\date{}

\begin{document}
\maketitle

\begin{abstract}
This paper describes the selection of a new southern/equatorial 
sample of Gigahertz Peaked Spectrum (GPS) radio galaxies, and 
subsequent optical CCD imaging and spectroscopic observations 
using the ESO 3.6m telescope. 
The sample consists of 49 sources with $-40^\circ<\delta<+15^\circ$,
$|b|>20^\circ$, and S$_{\rm{2.7GHz}}>0.5$ Jy, selected from the 
Parkes PKSCAT90 survey. About 80\% of the sources are optically 
identified, and about half of the identifications have 
available redshifts. The R-band Hubble diagram and evolution of 
the host galaxies of GPS sources are reviewed. 
\end{abstract}

\section{introduction}

Gigahertz Peaked Spectrum (GPS) radio sources are a class of powerful 
compact extra-galactic radio source characterised by an overall
turnover in their radio spectra at about 1 GHz in frequency
(for a review on GPS sources, see O'Dea 1998). 
They typically exhibit low flux density variability, implying that 
Doppler boosting is probably not important in these objects.
This view is strengthened by the VLBI morphologies -- of at least  
those identified with galaxies (not GPS quasars) -- which tend to show 
symmetric structures on both sides of a central core. These sources are 
therefore also called Compact Symmetric Objects (CSO). 
The overlap between the two classes is large, but may not be 
complete (see Snellen et al. 2000a for a discussion).

The characteristics of GPS sources are quite different from 
that of other classes of compact radio source, which can be
highly variable, and are almost always dominated by 
Doppler-boosted, one-sided core-jet emission. This may also be
the case for GPS sources optically identified with quasars which
also show core-jet VLBI morphologies. It has been argued that 
the GPS galaxies and quasars may not be related at all, but just 
happen to have similar radio spectra (Snellen et al. 1999).
In contrast to GPS galaxies, the GPS quasars are almost exclusively 
found at very high redshifts (O'Dea et al. 1991; Stanghellini et al. 
1998; de Vries et al. 1997).

CSO and GPS galaxies receive considerable attention, 
 because they are generally thought to be the young progenitors 
of large-size radio sources. Measurement of 
hot-spot advance speeds (Owsianik \& Conway 1998; Owsianik et al. 
1998; Taylor et al. 2000; Tschager et al. 2000) by multi-epoch VLBI 
observations have
proven for a handful of archetype GPS sources that these
objects are indeed young, typically 10$^{2-3}$ yrs.
In addition, measurements of the high frequency breaks in the 
slightly larger Compact Steep Spectrum (CSS) radio sources, indicate
 that they have radiative ages in the range of 10$^{3-5}$ yrs (Murgia et al. 1999).  
It is therefore generally believed that GPS galaxies evolve into
CSS and then FR\,I and/or FR\,II radio sources, assuming that they
are fuelled for long enough.
This makes CSO/GPS galaxies the key objects to study the early 
evolution of powerful radio-loud AGN (Fanti et al. 1995; Readhead et al. 1996;
O'Dea \& Baum 1997; Snellen et al. 2000b).

Complete samples of GPS sources are vital to perform statistical
studies. The complete sample of Stanghellini et al. (1998), 
containing sources with S$_{\rm{5GHz}}>1$ Jy, is the most 
extensively studied. Other samples include the 
faint GPS sources of Snellen et al. (1998), and that with GPS sources 
from the JVAS survey (Marecki et al., 1999) . Since radio telescopes
are mostly situated in the northern hemisphere, radio surveys are also
strongly biased towards the north. 
This hampers complete statistical studies 
of southern GPS sources, even at the highest flux density levels. 
Since the space density of GPS sources is so low, full sky coverage 
is important to firmly determine the behaviour of the high end of 
the luminosity function of GPS sources, which is vital 
for constraining radio source evolution models 
(Snellen et al. 2000b). In contrast, the most powerful optical
telescopes are found in the south (at least those available to European 
scientists), and therefore it is of great value to have a significant and
well defined sample
of GPS sources in the southern hemisphere.
Pioneering work on southern GPS sources has been done by de Vries et al. 
(1995; 2000),  especially on obtaining optical 
identifications and redshifts, and by King et al. (1996) involving
VLBI observations, but their samples
are not clearly defined. This paper describes the 
selection of a complete 
southern/equatorial sample of GPS {\it galaxies} from the 
Parkes PKSCAT90 database (Wright \& Otrupcek 1990), and presents the 
radio spectra and the optical identification status in a 
homogeneous way. Note that the GPS sources 
 identified as quasars are not included in the sample.
Section 2 describes the selection of the sample and its radio
properties. Section 3 describes optical CCD imaging and spectroscopy
performed with the ESO 3.6m telescope. In section 4, the 
results are presented and discussed, including an update on the 
R-band Hubble diagram for GPS galaxies.

\section{A southern/equatorial sample of GPS galaxies}
\subsection{Selection of the sample}
The basis for the selection of sources in the sample is the 
Parkes multi-frequency survey (PKSCAT90, Wright \& Otrupcek 1990), 
which 
is a compilation of flux density measurements at frequencies of 
178 MHz (4C), 408 MHz (Molonglo catalogue), and at 625, 1410, 2700,
5000, and 8400 MHz taken from the Parkes surveys. 
Firstly, all sources with $S_{\rm{2700MHz}}>0.5$ Jy were selected
in the regions with $-40^{\circ}<\rm{Dec}<+15^{\circ}$ and
galactic latitude, $|b|>20^{\circ}$ (1129 sources).
This flux density limit and galactic latitude cut-off were chosen
to be the same as those of the Parkes half-Jansky sample of
flat spectrum radio sources (Drinkwater 1997), allowing comparison of 
source statistics between the two samples. Note that 
most GPS galaxies are not found in the flat spectrum sample, 
since their high frequency spectrum is too steep.
Since 1410 MHz Parkes data is only sparsely available, the
positions of these sources were cross-correlated with the NVSS
survey (Condon 1998), of which all flux density was included within
a radius of 6$'$ from the Parkes position. In addition, the 
sources were cross-correlated with the Texas 365 MHz survey 
(Douglas et al 1996), which covers most of the sky area of the 
sample. Each spectrum was then fitted in an automated way 
with a 2nd order polynomial function  to find objects with a 
possible turnover. A total of 323 objects were pre-selected all of which
 showed a maximum above 100 MHz in the resulting fit. 
The spectra of these objects 
were then inspected visually to judge the validity of the fit.
The resulting 193 objects were cross-correlated with all Parkes 
radio sources in the NASA/IPAC Extra-galactic Database (NED)
with available redshifts, this to filter out any quasars. 
The NED database and
the CATS database of radio sources (Verkhodanov et al. 1997) were
then searched extensively for additional flux density points
and other information on the remaining 125 objects. In addition, 
the NVSS maps of each of the candidates were
checked to see whether or not any extended emission was present.
The low frequency survey measurements were taken
with interferometers, and therefore possible extended emission
may have been missed. 
This indeed turned out to be a major source of contamination for many
(47) mostly nearby radio galaxies. In other cases, the extra 
data points found in the NED and  CATS databases showed that the 
spectra did not turn over at frequencies below 300-400 MHz,
 or were subject to strong variability (19).
A few candidates turned out to be planetary nebulae (3), and 
some were BL Lacs or quasars without a measured redshift (7).
The spectra of all sources which were not pre-selected (806 objects)
were also visually inspected to see whether any candidates had been missed.
The remaining sample contains 49 radio sources with peaked spectra,
which are optically identified with a galaxy or still
 without an optical identification.

\subsection{The radio properties of the sample}

\begin{figure*}
\psfig{figure=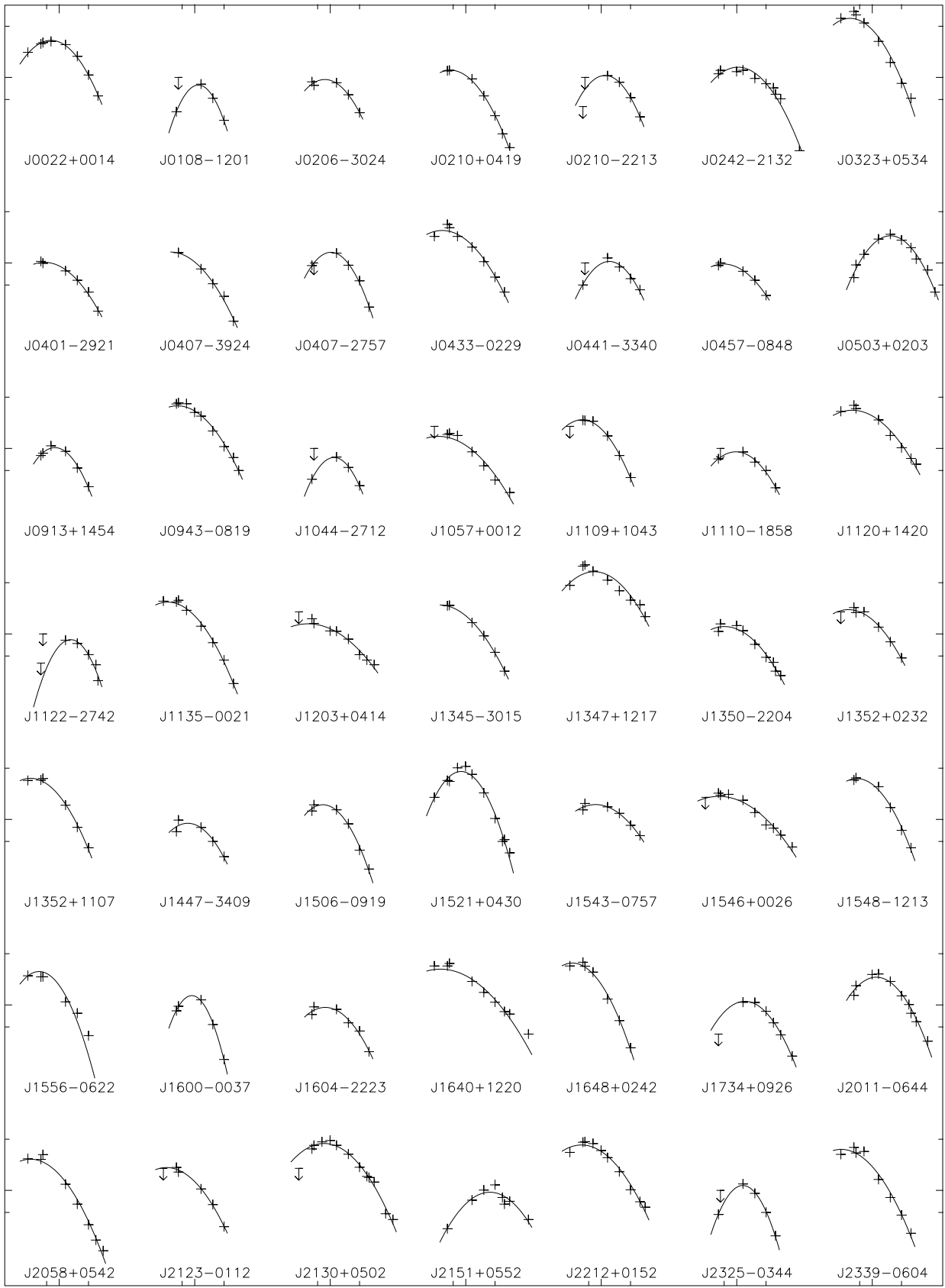,width=16.5cm}
\vskip 0.5cm
\caption{\label{radiospectra} The radio spectra of the GPS galaxies
in our sample. The tick-marks indicate 0.5, 1.0 and 5.0 GHz in 
frequency, and 0.5,1.0, and 5.0 Jy in flux density.}
\end{figure*}

From the 49 sources in the resulting sample, 12 are also present 
in the brighter samples of GPS sources from Stanghellini et al (1998)
and de Vries et al. (2000).
The properties of the objects are given in table \ref{radio}, 
with in column 1, the J2000 IAU name, in column 2 an alternative name, 
in column 3 the most accurate radio position found in the 
NED database or that from the NVSS survey 
(on which the optical identification process is based),
in column 4 the R-band magnitude (corrected to Cousins R and 
galactic extinction taken into account), in column 5
the redshift, in column 6 the flux density
at 2700 MHz, in column 7 the turnover frequency, in column 8
the flux density at the turnover, in column 9 the reference for the 
magnitude, in column 10 the reference for the redshift, and in column 11
a possible comment. The radio spectra
are shown in figure \ref{radiospectra}. The solid lines indicate
a 2nd order polynomial fit to the data-points to guide the eye,
and do not represent physically meaningful model-fits. The 
turnover frequencies given in column 6 of table \ref{radio}
are often just estimates, since it was not possible to determine
them accurately due to the sparse  sampling near their spectral peak.

\begin{table*}
\caption{\label{radio} The radio and optical properties of 
objects in the southern/equatorial sample of GPS galaxies.}
\begin{tabular}{ccccccccccccc}
IAU  &Other &Radio position&$m_{\rm{R}}$&z&$\rm{S}_{\rm{2.7GHz}}$&$\nu_{\rm{peak}}$&$\rm{S}_{\rm{peak}}$&Ref&Ref&comm.\\
Name &Name  & (J2000)      & mag& &       (Jy)       &    (GHz)          &  (Jy)     &$m_{\rm{R}}$&z\\
J0022$+$0014&4C$+$00.02 &00 22 25.48 $+$00 14 56.0&18.10$\pm$0.20&0.305&1.94&  0.6  & 3.1& 2& 2&A\\  
J0108$-$1201&B0105$-$122&01 08 13.20 $-$12 00 50.3&22.39$\pm$0.06&     &0.52&  1.0  & 0.9& 1&  & \\     
J0206$-$3024&B0204$-$306&02 06 43.26 $-$30 24 58.2&21.00$\pm$0.50&     &0.58&  0.5  & 0.9&13&  & \\
J0210$+$0419&B0208$+$040&02 10 44.52 $+$04 19 35.4&   $>$24.1    &     &0.56&  0.4  & 1.3& 1&  & \\     
J0210$-$2213&B0207$-$224&02 10 10.05 $-$22 13 36.6&23.52$\pm$0.13&     &0.86&  1.5  & 1.1& 1&  & \\       
J0242$-$2132&B0240$-$217&02 42 35.87 $-$21 32 26.2&17.10$\pm$0.50&0.314&0.97&  1.0  & 1.3&13&11& \\
J0323$+$0534&4C$+$05.14 &03 23 20.27 $+$05 34 11.9&19.20$\pm$0.50&     &1.60&  0.4  & 7.1&13&  & \\   
J0401$-$2921&B0359$-$294&04 01 21.50 $-$29 21 26.1&   $>$21.0    &     &0.58&  0.4  & 1.0&13&  & \\     
J0407$-$3924&B0405$-$395&04 07 34.43 $-$39 24 47.2&20.40$\pm$0.50&     &0.52&  0.4  & 1.4&13&  & \\   
J0407$-$2757&B0405$-$280&04 07 57.94 $-$27 57 05.1&21.14$\pm$0.04&     &0.93&  1.5  & 1.4& 1&  & \\
J0433$-$0229&4C$-$02.17 &04 33 54.90 $-$02 29 56.0&19.10$\pm$0.50&     &1.04&  0.4  & 3.0&13&  & \\   
J0441$-$3340&B0439$-$337&04 41 33.80 $-$33 40 03.6&   $>$21.0    &     &0.88&  1.5  & 1.2&13&  & \\  
J0457$-$0848&B0454$-$088&04 57 20.24 $-$08 49 05.2&20.30$\pm$0.50&     &0.58&  0.4  & 1.0&13&  & \\  
J0503$+$0203&B0500$+$019&05 03 21.20 $+$02 03 04.6&i21.0$\pm$0.20&0.583&2.46&  2.5  & 2.5& 4& 4&A\\
J0943$-$0819&B0941$-$080&09 43 36.86 $-$08 19 32.0&17.50$\pm$0.20&0.228&1.73&  0.4  & 4.2& 9&10&A\\ 
J0913$+$1454&B0910$+$151&09 13 35.00 $+$14 54 20.1&   $>$20.0    &     &0.54&  0.6  & 1.1&13&  & \\  
J1044$-$2712&B1042$-$269&10 44 37.63 $-$27 12 18.6&   $>$21.0    &     &0.55&  1.5  & 0.8&13&  & \\  
J1057$+$0012&B1054$+$004&10 57 15.78 $+$00 12 03.7&   $>$21.0    &     &0.58&  0.4  & 1.6&13&  & \\   
J1109$+$1043&B1107$+$109&11 09 46.04 $+$10 43 43.4&20.50$\pm$0.50&     &0.80&  0.5  & 2.4&13&  &E\\  
J1110$-$1858&B1107$-$187&11 10 00.45 $-$18 58 49.2&19.60$\pm$0.20&0.497&0.65&  1.0  & 0.9& 8&12& \\  
J1120$+$1420&4C$+$14.41 &11 20 27.81 $+$14 20 55.0&20.10$\pm$0.10&0.362&1.50&  0.4  & 3.7& 4& 4&A\\
J1122$-$2742&B1120$-$274&11 22 56.41 $-$27 42 48.2&   $>$21.0    &     &0.74&  1.4  & 0.8&13&  & \\  
J1135$-$0021&4C$-$00.45 &11 35 12.96 $-$00 21 19.5&16.50$\pm$0.50&     &0.76&  0.4  & 2.9&13&  &D\\   
J1203$+$0414&B1200$+$045&12 03 21.95 $+$04 14 17.7&18.80$\pm$0.50&     &0.52&  0.4  & 1.4&13&  & \\
J1345$-$3015&B1343$-$300&13 45 51.52 $-$30 15 04.1&   $>$21.0    &     &0.56&  0.4  & 2.5&13&  & \\
J1347$+$1217&4C$+$12.50 &13 47 33.36 $+$12 17 24.2&15.20$\pm$0.20&0.122&3.88&  0.4  & 8.8& 2& 2&A\\ 
J1350$-$2204&B1347$-$218&13 50 14.33 $-$22 04 43.7&20.93$\pm$0.05&     &0.72&  0.4  & 1.4& 1&  & \\  
J1352$+$0232&B1349$+$027&13 52 30.68 $+$02 32 47.7&20.00$\pm$0.50&     &0.78&  0.4  & 2.0&13&  & \\
J1352$+$1107&4C$+$11.46 &13 52 56.37 $+$11 07 07.7&   $>$21.0    &     &0.78&  0.4  & 3.6&13&  & \\   
J1447$-$3409&B1444$-$339&14 47 19.69 $-$34 09 16.2&21.00$\pm$0.10&     &0.50&  0.5  & 1.0&13&  & \\
J1506$-$0919&B1503$-$091&15 06 03.05 $-$09 19 12.5&19.70$\pm$0.50&     &0.87&  0.6  & 1.6&13&  & \\
J1521$+$0430&4C$+$04.51 &15 21 14.51 $+$04 30 20.0&22.10$\pm$0.11&1.296&2.30&  1.0  & 5.4& 6&10&A\\   
J1543$-$0757&B1540$-$077&15 43 01.69 $-$07 57 03.6&17.40$\pm$0.10&0.172&1.21&  0.4  & 1.7& 5& 5&C\\
J1546$+$0026&B1543$+$005&15 46 09.50 $+$00 26 24.6&20.10$\pm$0.20&0.556&1.24&  0.6  & 2.2& 2& 5&C\\
J1548$-$1213&B1545$-$120&15 48 12.97 $-$12 13 31.8&21.88$\pm$0.13&0.883&1.45&  0.4  & 3.7& 1& 1& \\  
J1556$-$0622&4C$-$06.43 &15 56 13.99 $-$06 22 37.8&22.20$\pm$0.13&     &0.77&  0.4  & 2.4& 1&  & \\
J1600$-$0037&B1557$-$004&16 00 00.91 $-$00 37 23.3&   $-$        &     &0.54&  1.0  & 1.2& 1&  &F\\    
J1604$-$2223&B1601$-$222&16 04 01.45 $-$22 23 41.3&18.75$\pm$0.10&0.141&0.57&  0.6  & 1.0& 5& 1& \\
J1640$+$1220&4C$+$12.60 &16 40 47.96 $+$12 20 02.1&21.36$\pm$0.20&1.150&1.48&  0.4  & 3.7& 1& 1& \\ 
J1648$+$0242&4C$+$02.43 &16 48 31.79 $+$02 42 46.0&  $>$21.0     &     &0.61&  0.4  & 3.4&13&  & \\   
J1734$+$0926&B1732$+$094&17 34 58.38 $+$09 26 57.8&20.80$\pm$0.10&     &1.08&  1.0  & 1.1& 5&  &B\\
%J2010$-$2425&B2007$-$245&20 10 45.13 $-$24 25 45.2& $>$21.0      &     &0.65&  0.4  & 2.0&13&  & \\
J2011$-$0644&B2008$-$068&20 11 14.22 $-$06 44 03.6&21.18$\pm$0.04&0.547&2.09&  1.4  & 2.6& 1& 1&A\\
J2058$+$0540&4C$+$05.78 &20 58 28.84 $+$05 42 50.7&23.40$\pm$0.30&1.381&0.65&  0.4  & 3.1& 7& 7& \\  
J2123$-$0112&B2121$-$014&21 23 39.12 $-$01 12 34.3&23.30$\pm$0.10&1.158&0.64&  0.4  & 2.0& 3& 2& \\ 
J2130$+$0502&B2128$+$048&21 30 32.88 $+$05 02 17.5&22.21$\pm$0.07&0.990&3.12&  1.0  & 4.8& 1&10&A\\ 
J2151$+$0552&B2149$+$056&21 51 37.88 $+$05 52 13.0&20.20$\pm$0.20&0.740&1.01&  5.0  & 1.2& 3&10&A\\
J2212$+$0152&4C$+$01.69 &22 12 37.97 $+$01 52 51.7&i22.0$\pm$0.20&     &1.80&  0.4  & 4.6& 4&  &A\\
J2325$-$0344&B2322$-$040&23 25 10.23 $-$03 44 46.7&23.50$\pm$0.20&     &0.91&  1.4  & 1.2& 9&  &B\\ 
J2339$-$0604&4C$-$06.76 &23 37 11.95 $-$06 04 12.4&22.91$\pm$0.20&     &0.80&  0.4  & 3.8& 1&  & \\   
\multicolumn{3}{l}{Comments:}\\
\multicolumn{10}{l}{A) Also in the sample of $>$1Jy GPS sources from Stanghellini et al. (1998); 
B) Also in the sample of de Vries et al. (1997)}\\
\multicolumn{10}{l}{C) Also in the sample of de Vries et al. (2000); D) Possibly a quasar; E) Large offset between the radio and optical position}\\
\multicolumn{10}{l}{F) Near a bright star, no magnitude}\\
\multicolumn{3}{l}{References:}\\
\multicolumn{10}{l}{1) This paper; 2) Snellen et al. (1996); 3) O'Dea et al. (1990); 4) de Vries et al. (1995); 5) de Vries et al. (2000)} \\
 \multicolumn{10}{l}{6) Biretta et al. (1985); 7) Stern et al. (1999); 8) Fugmann et al. (1988); 9) Stanghellini et al. (1993)}\\
\multicolumn{10}{l}{10) Stanghellini et al. (1998); 11) Wright \& Otrupcek (1990); 12) Drinkwater et al. (1995)}\\
\multicolumn{10}{l}{13) Digitized Sky Survey II; APM catalogue (Irwin et al. 1994); SuperCosmos Sky Surveys (Hambly et al. 2001)}\\ 
\end{tabular}
\end{table*}

\section{Optical ESO 3.6m Observations}

Optical CCD imaging and spectroscopy was performed on 
objects in the sample using the ESO 3.6m telescope at La Silla,
in Chile, on 27-28 July 2000, using the ESO Faint Object
Spectrograph and Camera (EFOSC2). The weather was photometric
throughout both nights. EFOSC2 uses a 2048$\times$2048 thinned 
Loral CCD chip with a 0.157 pixel size, which was binned by 
2$\times$2 for both the imaging and spectroscopy.

\subsection{CCD imaging}

Fourteen objects in the sample were observed in the $g$, $r$, 
and/or $i$ filters, which have effective wavelengths of 5169\AA, 
6814\AA, and 7931\AA$ $ respectively. For flux density calibration
standard star fields from Landolt (1992) were observed throughout the 
nights, and fluctuations in the magnitude scale zero-points were found 
to be smaller than 0.03 mag. The images were reduced in a standard way
in the software package IRAF from the NOAO, but were positionally 
calibrated within the AIPS software package (from the NRAO) using the 
task XTRAN. Positions of bright stars in the CCD fields were taken from 
the APM catalogue (Irwin et al. 1994) or the COSMOS UKST catalogue 
(Hambly et al. 2001), and in some cases directly measured on the 2nd Digitized 
Sky Survey. The flux calibrator measurements in Landolt (1992) were performed 
in the Johnson-Kron-Cousins $UBVRI$ photometric system, which is a 
somewhat different filter system from that used in this paper. 
The differences for the R and I band observations were found to be 
within the observational uncertainties, however for the $g$ band observations,
a colour correction to V-band had to be applied. 
From the observations of the standard star fields, this colour 
correction was found to be $V = g - 0.44 (V - R)$.
The magnitudes of the optical identifications were determined by adding
the flux within a box including all the emission from the object
nearest the radio position. In the case of multiple objects near 
the radio positions (possible merging or interacting galaxy systems),
the magnitudes were calculated for the total system and each component
individually. 

\begin{table}
\caption{\label{imaging} Details of the CCD imaging observations.}
\setlength{\tabcolsep}{0.75mm}
\begin{tabular}{rrcrcrc}\hline
Name & \multicolumn{2}{c}{V band} &\multicolumn{2}{c}{R band} &\multicolumn{2}{c}{I band}\\ 
     &  t & $m_V$          &  t & $m_R$ & t & $m_I$\\
     &  sec&   mag            &  sec& mag   &sec&mag \\   \hline 
J0108$-$1201& 60&23.50$\pm$0.23& 120&22.45$\pm$0.06&300&22.15$\pm$0.15\\
J0210$-$2213& 60&$>$24.40      &1500&23.55$\pm$0.13&300&23.68$\pm$0.32\\
J0210$+$0419&   &              &1500&$>$24.20      &   &              \\
J0407$-$2757& 60&21.94$\pm$0.19&  60&21.24$\pm$0.04&   &              \\
    GPS-west&   &22.61$\pm$0.16&    &21.71$\pm$0.11&   &              \\
        east&   &22.87$\pm$0.20&    &22.12$\pm$0.13&   &              \\
J1350$-$2204& 60&21.99$\pm$0.07& 300&21.13$\pm$0.05&300&20.30$\pm$0.12\\
   GPS-north&   &23.33$\pm$0.29&    &22.04$\pm$0.08&   &20.89$\pm$0.08\\
       south&   &22.59$\pm$0.12&    &21.93$\pm$0.10&   &21.50$\pm$0.13\\
J1548$-$1213& 60&22.93$\pm$0.17& 300&22.52$\pm$0.13&300&21.55$\pm$0.11\\
J1556$-$0622& 60&$>$23.40      &1500&22.82$\pm$0.13&1500&22.33$\pm$0.06\\
        east&   &              &    &23.40$\pm$0.15&   &22.93$\pm$0.13\\
    GPS-west&   &              &    &23.80$\pm$0.17&   &23.26$\pm$0.17\\
J1604$-$2223& 60&21.24$\pm$0.05&    &              &300&18.53$\pm$0.03\\
J1640$+$1220&   &              &  60&21.48$\pm$0.20&   &              \\
J2011$-$0644& 60&22.81$\pm$0.08&1500&21.52$\pm$0.04&300&20.48$\pm$0.06\\
       north&   &              &    &22.06$\pm$0.04&   &21.28$\pm$0.08\\
   GPS-south&   &              &    &22.48$\pm$0.07&   &20.99$\pm$0.07\\
J2130$+$0502&   &              &1500&22.38$\pm$0.07&750&21.05$\pm$0.03\\  
J2339$-$0604&   &              & 210&23.00$\pm$0.20&600&22.12$\pm$0.09\\\hline
                                     
\end{tabular}
\end{table}

\subsection{CCD spectroscopy}

We used the EFOSC2 grism\#06, which has 600 gratings per mm, resulting
in a dispersion of 2.06 \AA/pixel and a wavelengths coverage
of 3860-8070 \AA.A slit-width of 1$''$ was used resulting in a 
resolution of 13 \AA. Usually the slit was oriented near the 
paralactic angle. The reduction of the spectra was carried out in 
a standard way using NOAO's IRAF reduction software. 
Wavelength calibration was
carried out using a Helium-Argon lamp. One dimensional spectra 
were extracted by summing in the spatial direction over an aperture
as large as the spatial extent of the continuum or the brightest
emission line.

\begin{figure*}
\psfig{figure=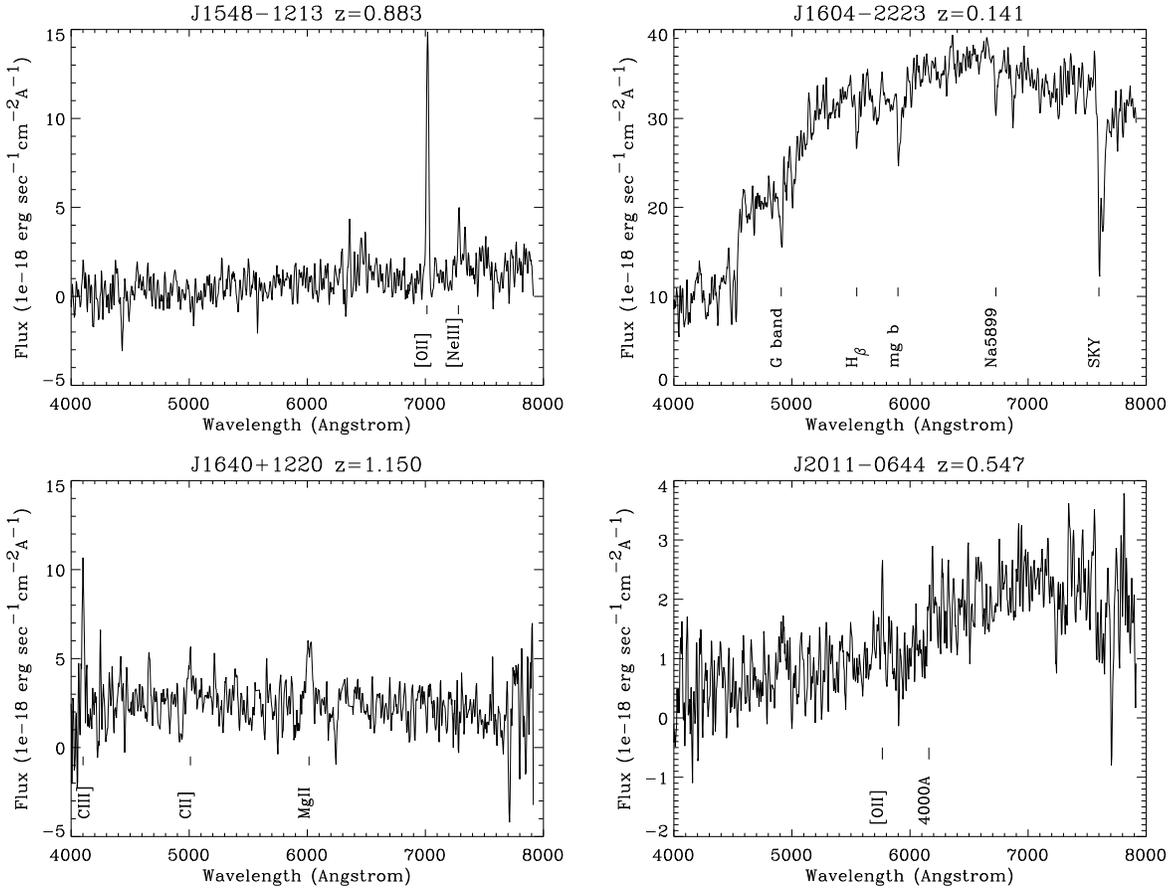,width=16cm}
\caption{\label{imspec} Optical spectra taken with EFOSC2 on the 
ESO3.6m telescope. Identified spectral features are indicated}
\end{figure*}

\begin{table}
\caption{\label{spectra} Details of the spectroscopic observations.}
\setlength{\tabcolsep}{0.75mm}
\begin{tabular}{lcrcl}\hline
Name&t&\multicolumn{2}{c}{Spectral feature}&z\\
    &sec&name& $\lambda_{\rm{obs}}$&\\
J1548-1213&1800&               &     &0.883$\pm$0.001\\
          &    &   [OII]3727\AA& 7019&  0.8833\\
          &    & [NeIII]3869\AA& 7285&  0.8829\\
J1604-2223&2700&               &     &0.141$\pm$0.001\\ 
          &    &       G4300\AA& 4909&  0.1416\\
          &    &     mgb5169\AA& 5898&  0.1410\\
          &    &      Na5899\AA& 6727&  0.1404\\
          &    &H$\beta$4861\AA& 5549&  0.1415\\
J1640+1220&4800&               &     &1.150$\pm$0.003\\
          &    &   MgII 2798\AA& 6015&  1.1497\\
          &    &    CII]2326\AA& 5010&  1.1539\\
          &    &   CIII]4101\AA& 1909&  1.1482 \\
J2011-0644&4200&               &     &0.547$\pm$0.001\\
          &    &   [OII]3727\AA& 5766&  0.5471\\
          &    &   break4000\AA& 6150&  0.54\\\hline
\end{tabular}
\end{table}

\subsection{Digitized Sky Surveys}
If the object was not observed by us, and no optical observation
was found in the literature, an image was extracted from the red
2nd Digitized Sky Survey. In addition, the 
APM catalogue (Irwin et al. 1992) and the SuperCosmos Sky Survey catalogues 
(Hambly et al. 2001) were searched for a possible identification and 
subsequently 
magnitude estimate. Unfortunately, the galaxy magnitudes in these
catalogues are only accurate to about 0.5 mag. DSS2 data were
retrieved for 22 sources (see fig \ref{dss}), with 9 objects 
remaining unidentified.

\section{Results and Discussion}

\begin{figure*}
\vspace{-0.5cm}
\hspace{-2.3cm}
\psfig{figure=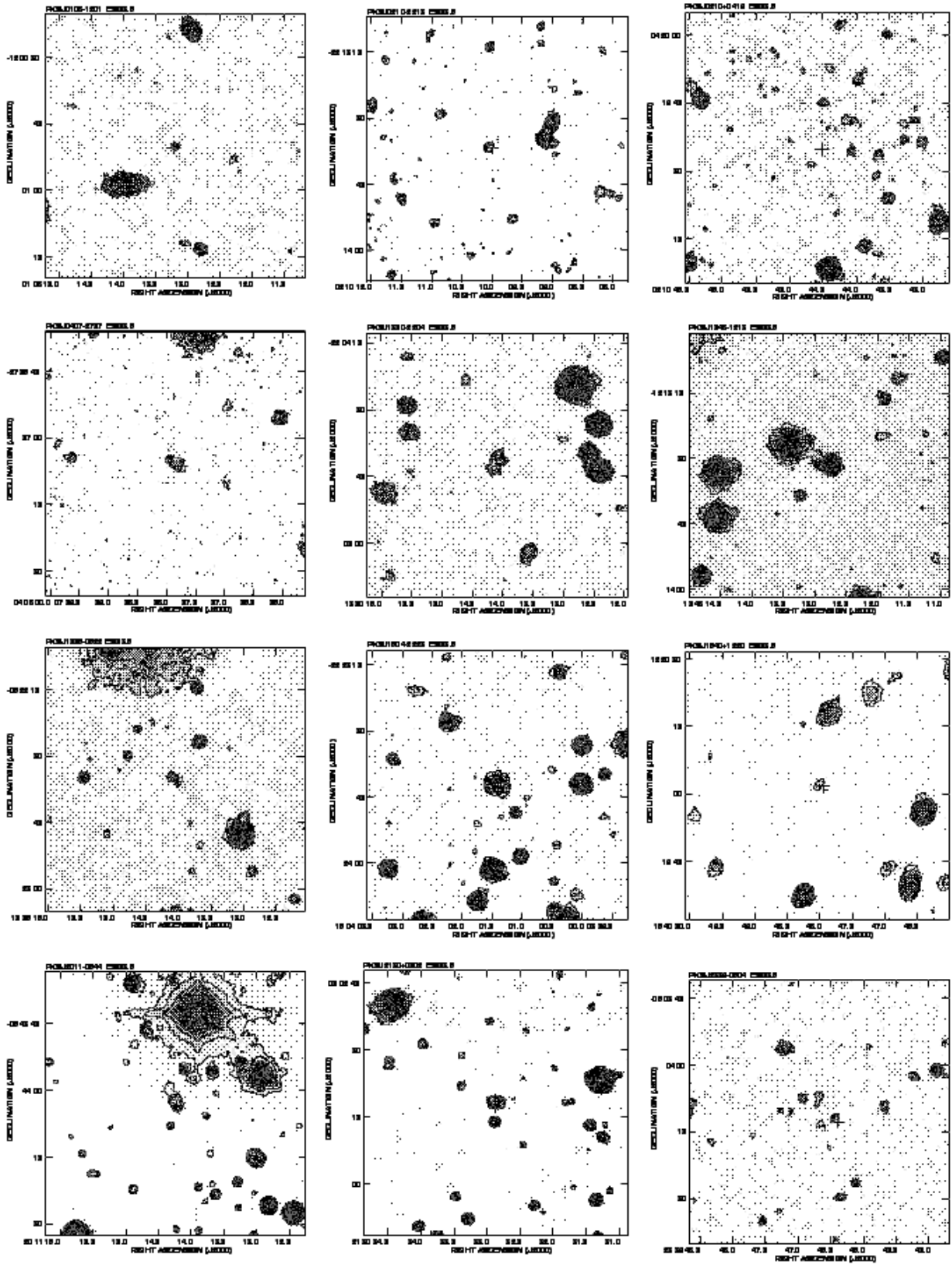,width=18cm}
\caption{\label{esofigs} Images from the ESO 3.6m observations.}
\end{figure*}

The photometric results of the imaging part of the EFOSC2 observations
are given in table \ref{imaging}, with in column 1 the source name,
in column 2 and 3 the exposure time and magnitude of the V band
observations, in columns 4 and 5 those of the observations in R band,
and in column 6 and 7 those of the I band observations.
The R band images are shown in figure \ref{esofigs}, except for 
J1604-2223, for which the I band image is shown.
Four objects (J0407-2757, J1350-2204, J1556-0622, J2011-0644) seem
to have a close companion object. For these objects, the total 
magnitude of the system is given, and those of the individual 
components, with that closest to the radio position indicated with 
'GPS' in column 1 of table \ref{imaging}. Note that the 
magnitudes in table \ref{imaging} have not yet been corrected for 
galactic extinction.

The results of the spectroscopic part of the EFOSC2 observations are
given in table \ref{spectra}, with in column 1 the source name, 
in column 2 the exposure time, and in column 3, 4, and 5 
the name, observed wavelength and derived redshift of the identified 
spectral features. Only the galaxies for which spectral features 
could be identified and subsequently their redshift determined,
are listed. Their spectra are shown in figure \ref{imspec}.
Spectra of several other galaxies were taken, but did not show any
identifiable features, and therefore no redshift could be determined.
Clearly these objects do not have emission lines of high equivalent width.

The ESO results and those of the DSS2 are incorporated in table
\ref{radio}, with in column 4 the R band magnitude, corrected 
for galactic extinction using the values of Schlegel et al. (1998).
The references to the optical magnitude and the redshift are 
given in column 9 and 10. The last column of table \ref{radio}
gives possible comments. One source, J1109+1043, 
has a large offset between the radio and optical position, 
casting doubt on its possible optical identification. The radio 
source J1600-0037 is located too close (17$''$) to a 12th 
magnitude star to allow an identification. This object is therefore
excluded from statistical analysis.
The optical identification of J1135-0021 could be point-like
and therefore may be a quasar. Strictly speaking it is not clear 
whether the empty fields are actually galaxies and not quasars.
However, from the magnitude distribution of other samples of 
GPS sources, it is rather unlikely that they are quasars
(most of the GPS quasars have m$_{\rm{R}}<21$). In general, only 
when all the sources are identified and spectra taken, can we be
confident that all the objects are galaxies.
The total identification fraction of the sample is $\sim$80\%.
About half of the identified objects have their redshifts
determined.

\subsection{The magnitude distribution}

\begin{figure}
\psfig{figure=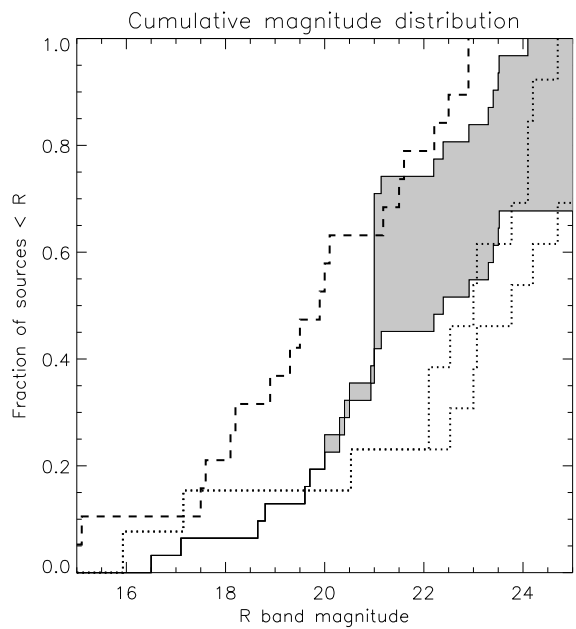,width=8.5cm}
\caption{\label{cummag} The cumulative R band magnitude distribution
of the half-Jansky sample of GPS galaxies, with the grey area 
indicating the uncertainty caused by the empty fields. 
The dashed line indicates the same distribution for the bright
1Jy Stanghellini et al. (1998) sample, and the dotted lines the 
limits to that of the faint GPS sample of Snellen et al. (1998).}  
\end{figure}

The apparent magnitude of a (GPS) galaxy is closely related to 
its redshift, as will be discussed later. Therefore, the magnitude 
distribution of a GPS galaxy sample is a reflection of 
its redshift distribution. This distribution is very interesting, since
it is closely related to the birth-function of radio galaxies
and the cosmological evolution of the GPS luminosity function
(Snellen et al 2000). The cumulative magnitude distribution
is shown in figure \ref{cummag}, with the shaded area indicating 
the distribution of the half-Jansky sample (taking into account 
the uncertainty of the magnitudes of the objects without optical 
identification). GPS sources with S$_{\rm{2.7GHz}}>1$ Jy were 
excluded to avoid overlap with the bright sample. 
The dashed line shows the magnitude distribution 
of the GPS galaxies of the $>$1Jy (at 5 GHz)  Stanghellini et al.
(1998) sample, which is completely identified, and the dotted lines
indicate the limits of the magnitude distribution of the galaxies 
in the faint GPS sample of Snellen et al. (1998). 
The mean flux density of the GPS sample of Stanghellini is about a 
factor of 4 brighter, and that of the faint GPS sample a factor
of 6 fainter than the half-Jansky sample (with S$_{\rm{2.7GHz}}<1$ Jy). 
Figure \ref{cummag} clearly shows that radio samples containing gernerally 
fainter radio sources are associated with galaxy samples with fainter 
magnitudes, with the median magnitudes of the 
bright, half-Jansky, and faint samples being about 20, 21, and 22
in R respectively. This seems to indicate that the radio-fainter
GPS galaxies are primarily at higher redshift, and not simply lower
in their radio luminosity. 

\subsection{The Hubble diagram, optical colours and absolute magnitudes}

\begin{figure}
\psfig{figure=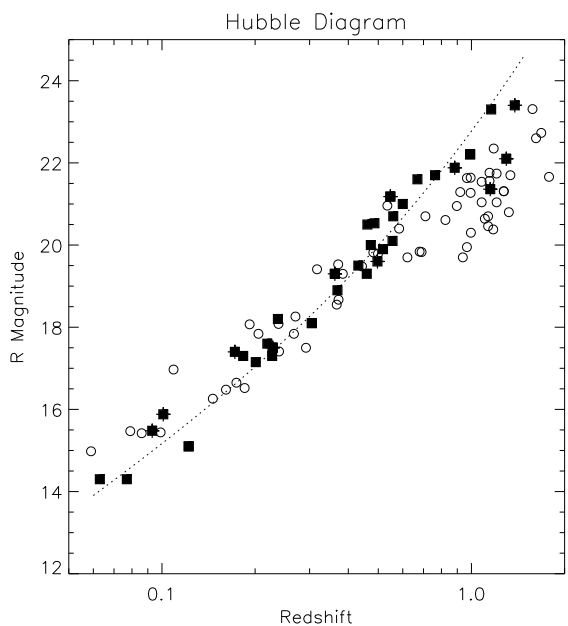,width=8.5cm}
\caption{\label{hubble} The Hubble diagram of GPS galaxies (solid squares) and 3C 
galaxies (open circles). The dotted line indicates
the Hubble relation for GPS galaxies as found by Snellen et al. (1996). 
The crossed GPS galaxies are those new to the sample.}
\end{figure}

The Hubble diagram of GPS galaxies has been previously discussed by 
Snellen et al. (1996a,1996b) and O'Dea et al. (1996).
Our group has claimed that the apparent magnitude - redshift 
relation of GPS galaxies show a low dispersion, and that they are apparently 
fainter at high redshift than 3C objects. This has been interpreted
in terms of  the GPS galaxies being intrinsically redder than 3C galaxies,
possibly due to the lack of the extra, blue light associated
with the radio-optical alignment effect (McCarthy et al. 1987; 
Chambers et al. 1987; Best et al. 1997). 
This view is strengthened by observations that the optical to near-infrared
colours of GPS galaxies are consistent with the light of these hosts being
dominated by old stellar populations (Snellen et al. 1996b; O'Dea et al. 1996).
In addition, Best et al. (1998) show that 
3C galaxies are typically up to a couple of magnitudes  bluer in R-K magnitudes
than expected for passively evolving elliptical models.
The lack of luminosity evolution 
found out to z=1, which would be expected for passively evolving ellipticals,
was interpreted as a possible dynamical evolution of the systems
through accretion, as found for brightest cluster galaxies 
(Aragon-Salamanca et al. 1998). Note that most faint (R$\sim$24) GPS galaxies
are found to have R-K$\sim$5-6 (Snellen et al. 1996b, 1998), 
and can therefore be classified as extremely red objects (ERO). 
Given their relative ease of selection, GPS galaxies promise to 
be excellent objects for studying the evolution of massive 
ellipticals out to $z>1$.

The combined samples of Stanghellini et al (1998), Snellen et al. 
(1998), and that presented in this paper provide an additional 15 GPS galaxies
 with available redshifts, compared to the original analysis of the 
R band Hubble diagram of Snellen et al. (1996). A few galaxies have 
only rough estimates of their R magnitude from the DSS surveys and 
are excluded from the analysis below, leaving 34 galaxies in total
(the galaxy J2151+0552 was excluded as well, since its optical light
is strongly influenced by non-thermal emission).
The updated Hubble diagram of GPS galaxies (solid squares) 
is shown in figure \ref{hubble}, in which all magnitudes are 
converted to Cousins R.
The dotted line indicates the Hubble relation as fitted to 
the original dataset in Snellen et al. (1996). The open circles 
indicate 3C galaxies from a compilation of samples from 
Best et al. (1997), Eales (1985), and de Koff et al. (1996) 
with, if necessary, the magnitudes converted to 'total magnitudes' in R Cousins.
As found before, the distant GPS galaxies are apparently fainter than
3C galaxies at similar redshift, which can be accounted for by the 
extra light of the alignment effect in 3C  galaxies (eg. Best et al. 1997).
However, some differences from the original
picture now emerge. On average, the nearby objects seem
to be fainter than in the original sample, and the galaxies at $z>1$ seem to 
be brighter than the original Hubble diagram indicates. 

\begin{figure}
\psfig{figure=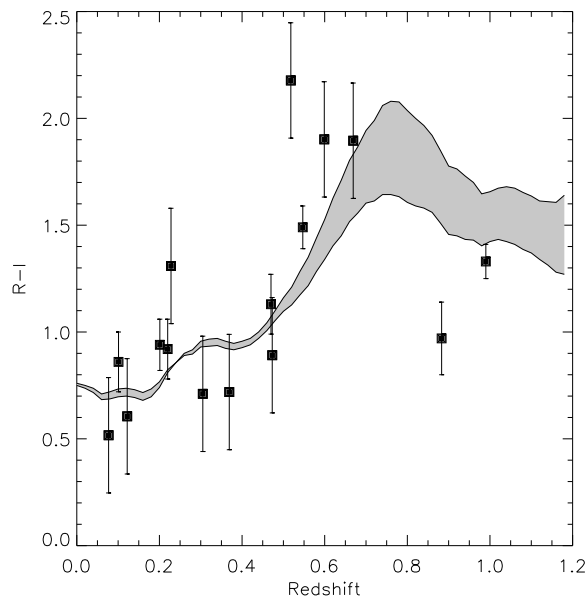,width=8.5cm}
\caption{\label{colour}The R-I colours as function of redshift for 
GPS galaxies taken from Stanghellini et al. (1993), Snellen et al. (1998) and this paper.}
\end{figure}

Figure \ref{colour} shows the R-I colours as function 
of redshift, with sources taken from Stanghellini et al (1993),
Snellen et al. (1999) and this paper (all corrected to Cousins R).
The shaded area indicates the expected R-I colours for 
a passively evolving elliptical with a formation redshift of z=5,
and a range from 1 to 2 times solar metallicity 
(Worthey et al. 1994). The R-I colours 
are more or less as expected for passively evolving ellipticals, 
as shown before by O'Dea et al. (1996). 

\begin{figure}
\psfig{figure=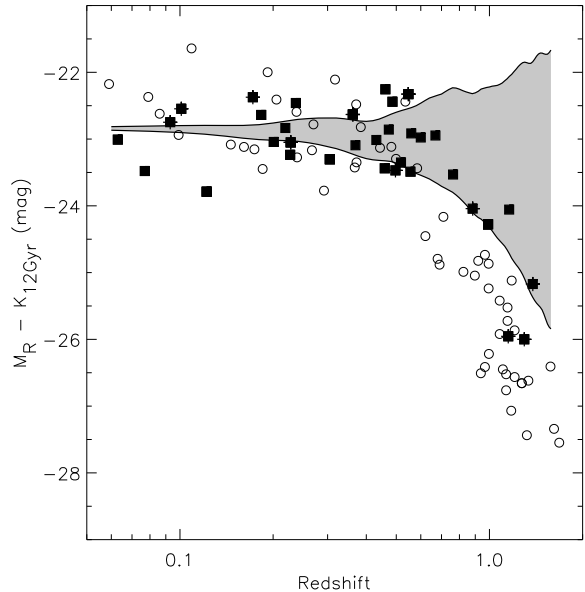,width=8.5cm}
\caption{\label{absmag} The absolute magnitudes of the host galaxies
of GPS sources (solid squares) and 3C galaxies (open circles) as 
function of redshift. The magnitudes are k-corrected using a 
synthetic spectrum of a 12 Gyr old elliptical with solar metallicity
from Worthey (1994). The crossed GPS galaxies are those new to the 
sample.}
\end{figure}

Figure \ref{absmag} presents the absolute magnitudes for GPS galaxies 
(solid squares) and 3C galaxies (open circles). A cosmology
with $H_{\rm{0}}$=70 km sec$^{-1}$Mpc$^{-1}$, $\Omega_{\lambda}=0.3$ and 
$\Omega_m=0.7$ is assumed, and  the magnitudes are K-corrected using a 
synthetic galaxy spectrum by Worthey (1994) for a 12 Gyr old elliptical
with solar metallicity. The models of Worthey (1994) are also
used to calculate the grey area, which represents the expected
luminosity evolution as function of redshift for a 2$L_*$ 
elliptical with a formation redshift of z=5, and a range of 
metallicities from 1 to 2 times solar. 

As for the Hubble diagram, this picture seems to be different
to that found earlier for the absolute K-band magnitudes of 
GPS galaxies as function of redshift (Snellen et al. 1996b), 
which show a more or less constant absolute magnitude as 
function of redshift. We believe this has several reasons.
Firstly, the few low redshift GPS galaxies available in 
1996 seem to have been, on average, over-luminous with respect to the 
general low redshift population. This was already a concern at that time, 
since two of them have a Seyfert nucleus (4C+12.45 and 
B1404+286). This has moved the z=0 baseline to a lower luminosity,
making the higher redshift objects relatively brighter.
Secondly, the now popular $\lambda$-cosmology makes the objects
at redshift of unity intrinsically about half a magnitude brighter 
as well.
Thirdly, the newly measured redshifts at $z>1$ are strongly biased
towards the objects with the strongest emission lines, making 
contamination from an active nucleus more likely. It is therefore 
expected that the majority of the objects at $z>1$, for which we were
not able to determine a redshift, are intrinsically less luminous
than the few objects shown in figures \ref{hubble} and \ref{absmag}.
Indeed, from the four spectra shown in figure \ref{imspec},
the emission lines of the two galaxies at low redshift 
have very low equivalent widths, while those of 
the two at high redshift have very high equivalent width. 
For that reason only the redshifts of the latter two could be
determined. Several deep spectra of other faint galaxies 
were taken with no lines detected. Eight-metre-class telescope
time will be needed to resolve this issue.

\subsection{Mergers and interactions}

Previous work has indicated that disturbed optical 
morphologies and the presence of close companions are common
characteristics of GPS galaxies (Stanghellini et al. 1993,
O'Dea et al. 1996). Using a sample of radio-bright GPS galaxies, O'Dea et al. 
find that about half of the galaxies have disturbed
morphologies and about one-fifth show a 2nd component within
a few arcseconds. Very similar percentages have been found by
Heckman et al. (1986) for a general sample of powerful radio-loud AGN.
If GPS sources evolve into large scale radio sources, then it would
be expected that their galactic environments are similar, since
the timescales of interaction and/or mergers are larger than that 
of the lifetime of a radio source.

The statistics in the sample presented here
are not yet good enough to perform a similar statistical analysis.
However, it is interesting to note that from the 12 galaxies
we imaged, 4 are clearly part of a multiple system
(J0407-3924, J1350-2204, J1556+0622, and J2011-0644). In addition,
J1548-1213 seems to have a 2nd component, but the picture is less
clear due to 2 foreground stars nearby. At first glance, this 
gives a similar fraction of multiple systems to O'Dea et al. (1996)
and Heckman et al. (1986). It is also interesting to note that in all 4 
of the multiple systems, the GPS radio sources are identified with the 
redder component of the two. A more extensive observing programme
for optical imaging and spectroscopy is planned. 

\section{Summary and Conclusions}

In this paper we have described the selection of a new southern/equatorial sample of Gigahertz Peaked Spectrum (GPS) radio galaxies,
and presented subsequent optical CCD imaging and spectroscopic 
observations using the ESO3.6m telescope. 
The sample consists of 49 sources with $-40^\circ<\delta<+15^\circ$,
$|b|>20^\circ$, and S$_{\rm{2.7GHz}}>0.5$ Jy, selected from the 
Parkes PKSCAT90 survey. About 80\% of the sources are now optically 
identified, and about half of the identifications have 
their redshifts determined. The R-band Hubble diagram and evolution of 
the host galaxies of GPS sources are reviewed. The view on GPS galaxies 
seems to be slightly altered, with the host galaxies increasing in optical 
luminosity with redshift, as expected for passive evolution. 
This change appears to be caused by the low-z 
galaxies being less luminous than previously thought, 
the new $\lambda$-cosmology used, and a strong observational bias towards high 
redshift GPS galaxies being those with strong emission lines.

The sample presented in this paper has the potential to 
greatly improve the statistics on luminous young radio-loud AGN,
and the use of hosts of GPS sources for 
galaxy evolution studies, especially with the VLT and Gemini-south
telescopes. 

\section*{Acknowledgements}
This research has made use of observations collected at the ESO/La Silla 3.6m telescope, and of the NASA/IPAC Extragalactic Database 
(NED) which is operated by the Jet Propulsion Laboratory, California
 Institute of Technology, under contract with the National 
Aeronautics and Space Administration. 
We made use of the CATS database (Verkhodanov et al. 1997) of the Special 
Astrophysical Observatory.
The Second Palomar Observatory Sky Survey (POSS-II) was made by the California 
Institute of Technology with funds from the National Science Foundation, 
the National Aeronautics and Space Administration, the National Geographic 
Society, the Sloan Foundation, the Samuel Oschin Foundation, and the Eastman 
Kodak Corporation. 
{}

\section*{Appendix}

\begin{figure*}
\vspace{-0.0cm}
\hspace{-1.3cm}
\psfig{figure=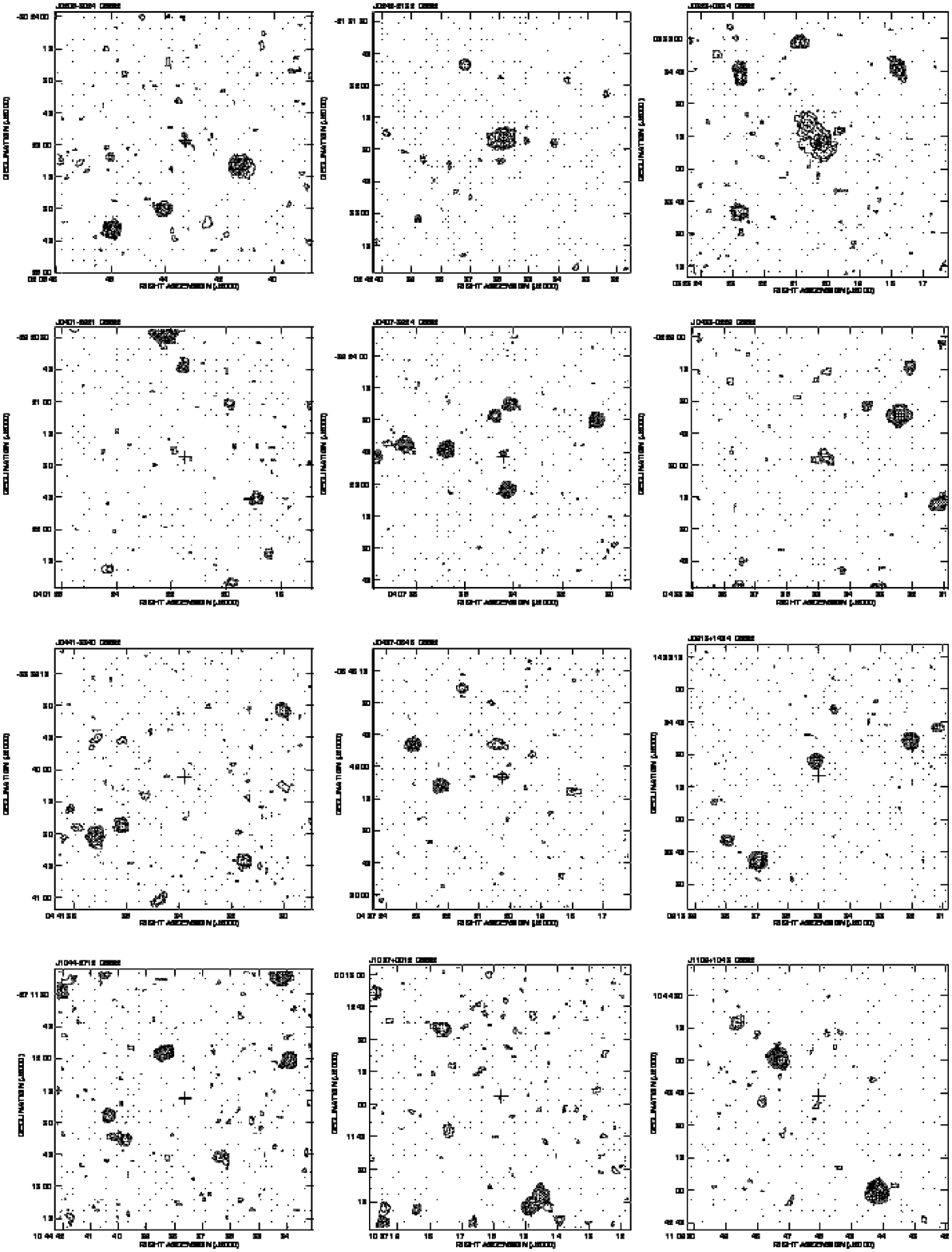,width=18cm}
%\vspace{-1.5cm}
\caption{\label{dss} Images retrieved from the red 2nd Digitized Sky Survey}
\end{figure*}
\addtocounter{figure}{-1}
\begin{figure*}
\vspace{-0.0cm}
\hspace{-1.3cm}
\psfig{figure=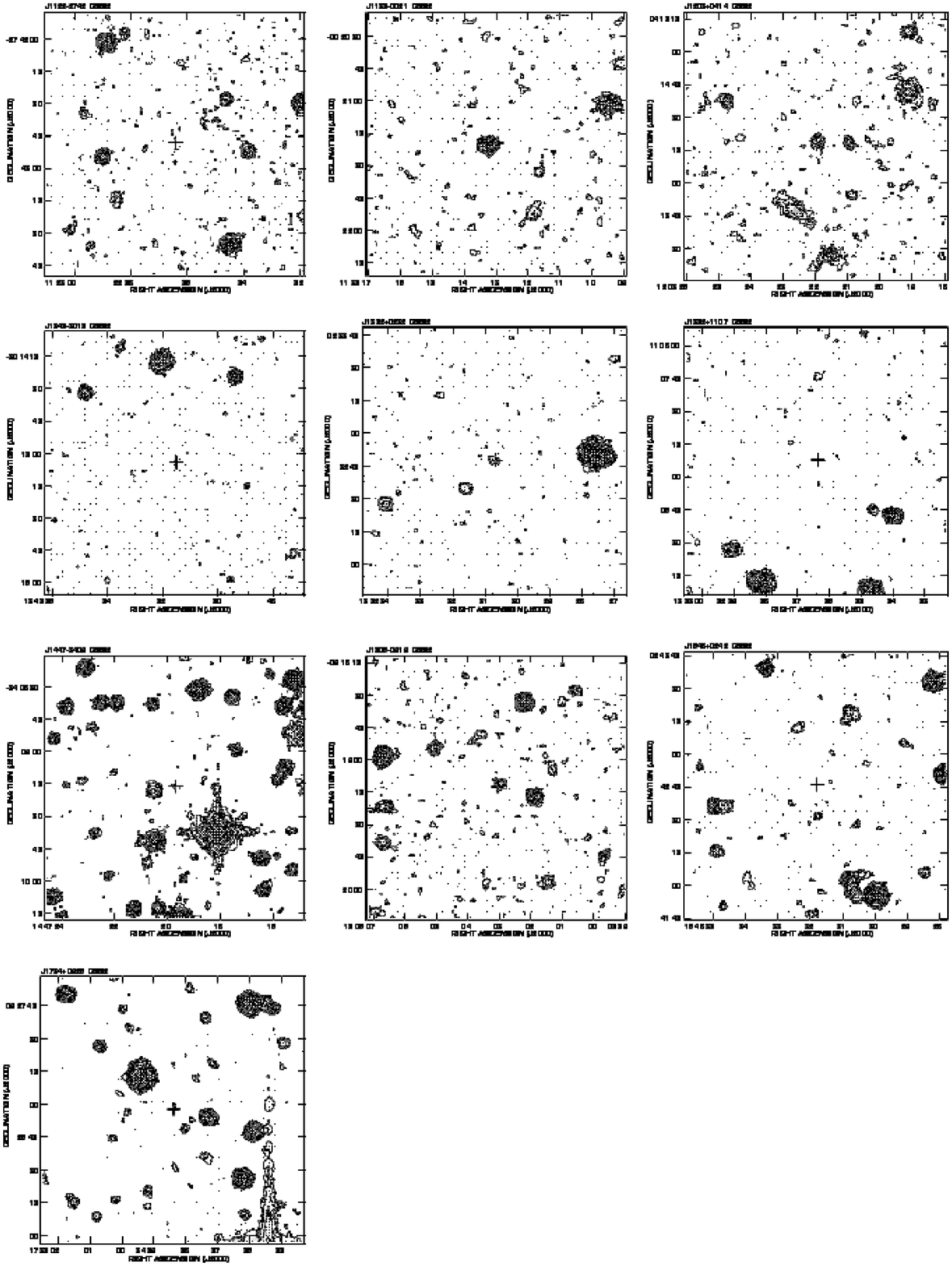,width=18cm}
%\vspace{-1.5cm}
\caption{\label{dss} {\it continued....}}
\end{figure*}

\end{document}